# A MICROFLUIDIC PORE NETWORK APPROACH TO INVESTIGATE WATER TRANSPORT IN FUEL CELL POROUS TRANSPORT LAYERS

**A. Bazylak, V. Berejnov, B. Markicevic, D. Sinton, and N. Djilali**
Email: abazylak@uvic.ca (Bazylak), berejnov@uvic.ca (Berejnov), bmarkice@uvic.ca (Markicevic),
dsinton@engr.uvic.ca (Sinton), ndjilali@uvic.ca (Djilali)
Dept. of Mechanical Engineering and Institute for Integrated Energy Systems,
University of Victoria, Victoria, British Columbia, Canada

## ABSTRACT
Pore network modelling has traditionally been used to study displacement processes in idealized porous media related to geological flows, with applications ranging from groundwater hydrology to enhanced oil recovery. Very recently, pore network modelling has been applied to model the gas diffusion layer (GDL) of a polymer electrolyte membrane (PEM) fuel cell. Discrete pore network models have the potential to elucidate transport phenomena in the GDL with high computational efficiency, in contrast to continuum or molecular dynamics modelling that require extensive computational resources. However, the challenge in studying the GDL with pore network modelling lies in defining the network parameters that accurately describe the porous media as well as the conditions of fluid invasion that represent realistic transport processes. In this work, we discuss the first stage of developing and validating a GDL-representative pore network model. We begin with a two-dimensional pore network model with a single mobile phase invading a hydrophobic media, whereby the slow capillary dominated flow process follows invasion percolation. Pore network geometries are designed, and transparent hydrophobic microfluidic networks are fabricated from silicon elastomer PDMS using a soft lithography technique. These microfluidic networks are designed to have channel size distributions and wettability properties of typical GDL materials. Comparisons between the numerical and experimental flow patterns show reasonable agreement. Furthermore, the fractal dimension and saturation are measured during invasion, revealing different operating regimes that can be applied to GDL operation. Future work for model development will also be discussed.

## INTRODUCTION
Despite the early development of pore networks by Fatt in 1956 (Fatt, 1956; Fatt, 1956; Fatt, 1956) and Broadbent and Hammersley in 1957 (Broadbent & Hammersley, 1957), it was only in the past two decades that pore network models became heavily employed for studying displacement processes in idealized porous media (Wilkinson and Willemsen, 1983). Since this revival of pore network modelling, pore networks have been applied to areas ranging from oil recovery (Chandler et al., 1982; Payatakes, 1982; Koplik and Lasseter, 1985; Dixit et al., 1999; Valavanides and Payatakes, 2001; Paterson, 2002), ground water hydrology (Bernadiner, 1998), textile characterization (Thompson, 2002), and most recently to gas diffusion electrodes for the polymer electrolyte membrane (PEM) fuel cell (Gostick et al., 2007; Markicevic et al., 2007; Sinha et al., 2007).

The gas diffusion electrode commonly referred to as the gas diffusion layer (GDL) or gas diffusion media is typically a 200 µm layer of the PEM fuel cell composed of carbon fibers either oriented randomly and pressed together (carbon paper) or woven together (carbon cloth). The GDL is sandwiched on both sides of the membrane electrode assembly, as shown in Figure 1 (a), and provides several key functions. It provides mechanical support for the membrane electrode assembly. The highly conductive carbon fibers provide pathways for electronic transport between the catalyst layer and the current collecting plates. The void spaces that result from the fiber orientation provide pathways for gaseous fuel transport and excess heat removal. The GDL is also treated with a hydrophobic coating to enhance excess liquid water removal. Excess liquid water is a major limiting factor to increased performance, durability and lifetime, which is why there is great interest to investigate liquid water transport in this complex hydrophobic media. However, the opaque nature of the GDL, as shown in Figure 1 (c), currently used in the PEM fuel cell greatly limits the direct observation of through-plane water transport. This limitation provides challenges to identifying and understanding the transport processes that govern two phase flow in this complex media, and in turn this inhibits the advancement of water management techniques needed to improve fuel cell performance.

Pore network models provide an alternative to continuum phase models to elucidate transport phenomena and determine transport parameters in porous media. Continuum modelling of porous media becomes unattractive when the porous media has a complex three-dimensional heterogeneous and disordered geometry, such as the gas diffusion layer. Numerically tracking phase interfaces and calculating surface tension forces becomes computationally expensive. In contrast, pore network modelling is an intermediate approach where the disordered porous media is represented by an equivalent network of pores and throats, and the throats are assigned varying hydraulic conductivities.

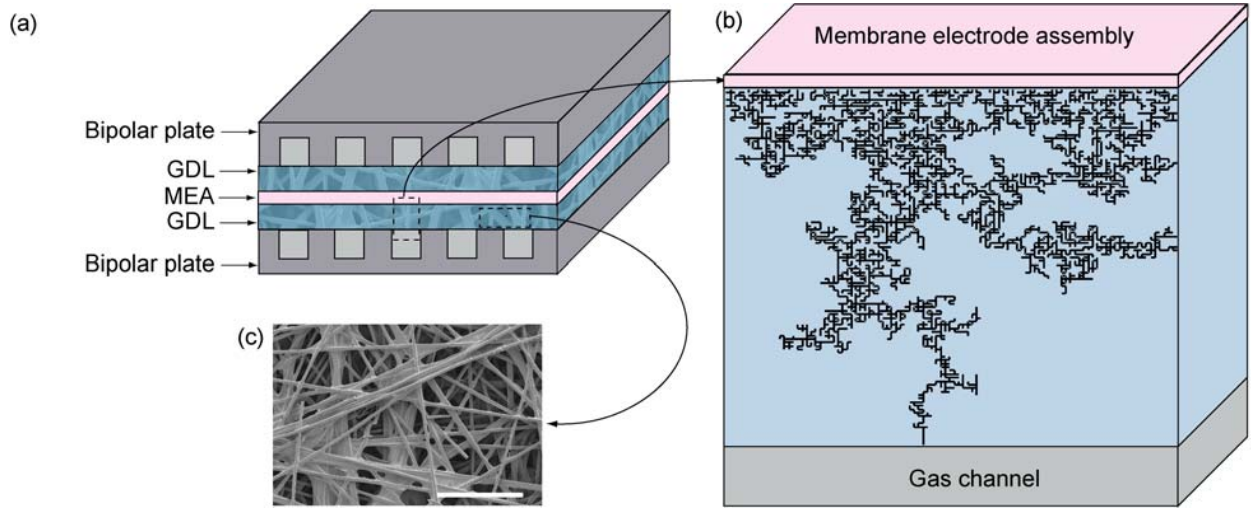

**Figure 1.** Schematic showing pore network modelling results with respect to PEMFC geometry: (a) diagram of a PEMFC, (b) sample numerical pore network breakthrough pattern of a simulated gas diffusion layer, and (c) scanning electron microscope image of Toray TGP-H-060 GDL (length bar represents 200 µm).

In terms of PEMFC modelling, pore network models are complimentary to continuum modeling, where complex two-phase microscale flows are reduced to constitutive relationships that are required as inputs to the continuum models.

In this work, we discuss developing and validating a GDL-representative pore network model. The first stage consists of creating a pore network geometry with pore sizes characteristic of the GDL (Mathias et al., 2003). Although in reality, the GDL is a fibrous substrate with pore volumes resulting from the intersection of cylindrical fibers, by effectively mapping the physical properties of the GDL (pore size and spatial distribution, wettability, etc.) the pore network model provides a valuable tool for predicting the constitutive relationships that are required for continuum phase models. An in-house software package (Markicevic et al., 2007) featuring invasion percolation with trapping (IPT) is employed to numerically model the drainage process (Wilkinson and Willemsen, 1983). Transparent experimental microfluidic chips containing these pore networks are employed to compliment the development of a GDL-representative pore network model. Invasion flow patterns are visualized in the transparent microfluidic networks using fluorescence microscopy, and the time series evolution of liquid water transport through the networks are measured. In keeping with the random nature of GDL materials, all pore networks studied here are generated with a random distribution of throat widths. Longer range geometrical variations (biasing) added to these networks are of interest in comparing the influence on water transport.

Although the 2D pore network approach does have limitations. For instance, 2D and 3D simulations have different connectivities and yield different percolation thresholds. Nevertheless, 2D network systems are expected to be qualitatively representative, which allow acceptable simulation turnaround times for the first stage of pore network model development

## A MICROFLUIDIC PORE NETWORK APPROACH

### Structure of the Pore Network Pattern

The pore network generation procedure begins with a regular lattice of square obstacles. Figure 2 (a) shows a schematic of the regular lattice, where $w$ is the throat width, and $L$ is the obstacle square side length. A random translational perturbation is then applied to each obstacle, where $\Delta X$ refers to the horizontal translation, and $\Delta Y$ refers to the vertical translation of the original obstacle. The uniform random distribution of translational perturbations is shown in Figure 2 (d). Figure 2 (c) is the resulting schematic of a portion of the random pore network. For a uniform channel height of 25 µm, the hydraulic throat radii distribution is shown in Figure 2 (e). All hydraulic throat radii employed in this work are characteristic of Toray TGP-H-060 gas diffusion layers (Mathias et al., 2003).

To generate an isotropic random pore network, the translation of the regular lattice in the horizontal direction $\Delta X$ and the vertical direction $\Delta Y$ must be independent. It is also possible to generate a random pore network with a diagonal bias by coupling the translation of the regular lattice in the horizontal and vertical directions. To implement a diagonal bias of $\theta_{bias} = 45°$, for each square obstacle, $\Delta Y = \Delta X$. The ratio of vertical to horizontal obstacle translations can be adjusted to implement the desired diagonal bias for $0° < \theta_{bias} < 90°$, as shown in Eqn. (1).

$$\tan \theta_{bias} = \frac{\Delta Y}{\Delta X} \qquad (1)$$

For example, equalizing ΔX and ΔY translations produces a pattern with the appearance of a "diagonal" bias directed at 45 degrees.

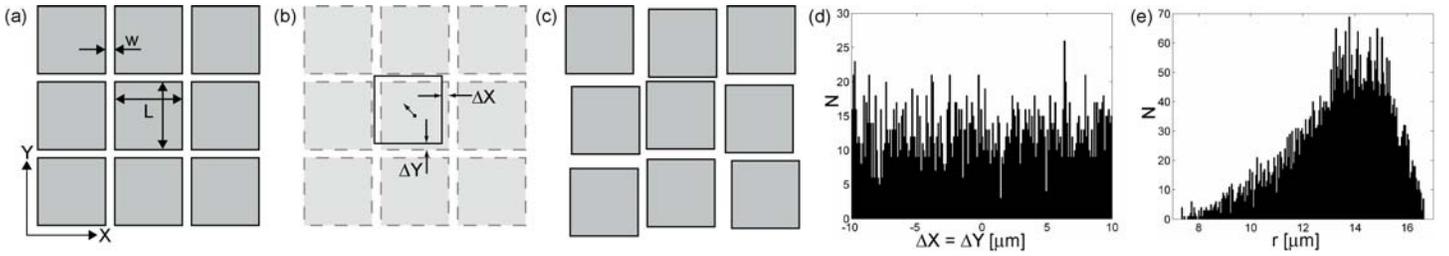

**Figure 2.** Schematic illustrating the photomask generation process: (a) regular square lattice, (b) translation of regular lattice in both the horizontal and vertical directions, (c) resulting arrangement of random pore network, (d) uniform distribution of horizontal and vertical translations, and (f) hydraulic throat radius distribution.

An invasion percolation algorithm with trapping was used to investigate deterministic realizations of pore network filling. The two-dimensional model accounted for capillary dominated drainage and cluster formation in a regular square capillary network of size $n_L$ x $n_L$ pores. The porous media was represented by a matrix of pores and connecting throats. The invading phase entered the network along the top side, and the invading fluid propagated through the throat with the largest throat radius through a series of discrete quasistatic steps. The fluid exited through the opposite side to the entrance, and the invasion was stopped once breakthrough was reached, where breakthrough is defined as when the first throat at the outlet is filled with the invading fluid. In this work, the deterministic nature of pore network invasion provided a useful tool to compare with experimental results. Figure 1 (b) shows a sample numerical breakthrough pattern in relation to a fuel cell schematic.

**Microfluidic Fabrication**

The microfluidic networks employed in this work were fabricated with a conventional soft-lithography technique (Xia and Whitesides 1998; McDonald, Duffy et al. 2000). Some modifications were made to the conventional technique in order to build large networks with variable material surface properties, as shown in Figure 3. The fabrication of the microfluidic network consisted of four stages: designing and printing a network photomask (Figure 3 (a)), fabricating a master (Figure 3 (b, c)), fabricating an elastomer replica (Figure 3 (d)), and assembling the parts into a microfluidic chip (Figure 3 (e1)). The final device was a transparent elastomeric microfluidic chip, which consisted of a system of rectangular (25 x $L$ x $W$ μm) microscale channels embossed in a PDMS polymer (Figure 3 (e)). The length, $L$, and width, $W$, of the channels were distributed in the following ranges: $L \in$(200-300 μm) $W \in$(10-130 μm), respectively. The networks were square with overall dimensions approximately 15x15 mm and with approximately 5000 throats. When fabricating microfluidic networks, the challenge is to quickly prepare a large number of representative network patterns with high pore/throat density that regularly deviate from one another.

As shown in Figure 3 (a), the mask manifold consisted of: delivery channels (1, 2), pressure-drop channels (3), inlet/outlet manifolds (4), and the pore network pattern. Once the network patterns were designed and generated, they were inserted into the AutoCAD template. AutoCAD drawings were printed on transparencies with a high resolution (5K, 10K, and 20K dpi) commercial printer (Outputcity, California). These transparences were employed as photomasks for fabricating the negative masters.

A microscope glass slide (75x51 mm, Esco, NH) or silicon wafer (3" Silicon Quest International Inc.) was spin-coated with SU8-25 epoxy negative tone photoresist (MicroChem, USA) and subjected to a pre-baking period ~ 20 min. A photomask was placed over the solid photoresist film and exposed to UV light from a mercury lamp (Tamarack Scientific PRX 200/350 UV Collimated Exposure System) as shown in Figure 3 (b). Approximately four minutes of UV light exposure were required to successfully develop the narrowed throat (15 μm). After the exposure, the slide was post baked for approximately 15-20 min at 95°C to allow the cross linking polymer reaction to complete. The slide was submerged in SU8 developer (MicroChem, USA) for approximately 2 min to remove the non-polymerized SU8-25 photoresist. The developed slide was dried by a high pressure air jet and heated at 95°C for 30 min to yield the hard negative relief of the channel network.

A polymethylsiloxane (PDMS) polymer (Sylgard 184, Dow Corning, NY) polymer was cast against the master, which was placed in the plastic Fisherbrand Petri dish, as shown in Figure (d). Prior to casting, the mixture of PDMS prepolymer and curing agent (10:1) was degassed to eliminate bubbles. The PDMS was cured, and the replica was cut with a blade and peeled away from the master. This yielded an elastomeric replica containing a positive structure of the network channels. The inlet and outlet ports were punched with rounded holes to hold plastic capillary tubing (1/16[th] inch OD Teflon tubing, Fisher Scientific). The thickness of all replicas was approximately 5 mm, which enabled firm connections with chip ports without additional inlet/outlet connectors.

The PDMS replica was combined with a PDMS covered glass slide, as shown in Figure 3 (e1). The PDMS layer on the glass provided a flat and thin substrate for the microfluidic network. We used both irreversible (Figure 3 (e1)) and reversible (not shown) assembling methods in our experiments. Irreversible assembling (Duffy, McDonald et al. 1998) is based on the oxygen plasma treatment of PDMS. This method permanently seals PDMS surfaces to prevent inter-pore and inlet connector/delivery channel leakages. In the reversible assembly method, the elastic replica and PDMS covered substrate were compressed with clamps to provide a firm contact and prevent leakages. With this reversible seal, the chip was opened, cleaned, and treated for multiple uses. In terms of the numerical simulations, the single phase gas permeability only varied by 3.2% ($10^{-14}$ m$^2$) when the random pore networks were un-biased and diagonally biased. Although the pore network model does not have a definable porosity, the calculated porosity of the experimental pore network ranged between 0.3 and 0.6.

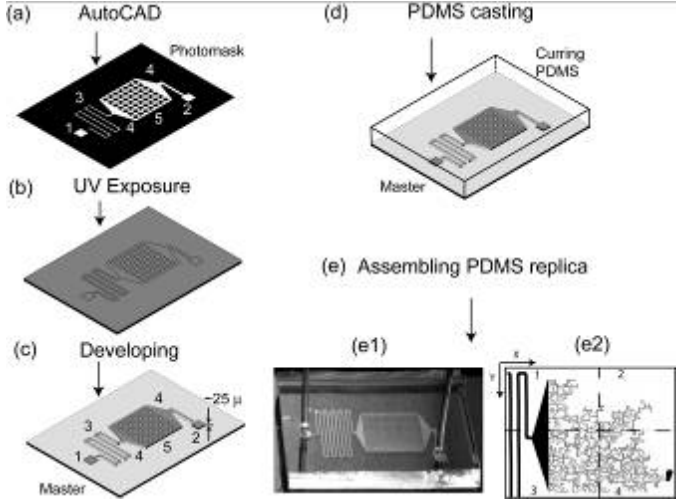

**Figure 3.** Fabrication of the microfluidic network chips. (a) a high resolution network photomask is produced using AutoCAD; (b) a flat substrate coated with a thin photoresist layer of thickness ~ 25µm, and UV exposure causes photoresist to polymerize; (c) non-polymerized parts of photoresist are chemically etched to leaving a planar 3D network structure (master); (d) a PDMS cast produces the elastomeric replica, which consists of the network pattern; the network chip can be assembled irreversibly within the PDMS base (e1) and the flow pattern can be measured (e2).

**Flow Visualization**

Flow visualization and network scanning was achieved using a Leica DMI 6000B fluorescence inverted microscope: HCX PL Fluotar 1.25x0.04 objective (Leica, Germany) with Leica LAS AF fluorescence-image-operation software. The microscope stage scans were synchronized with image acquisition using a high resolution, highly sensitive CCD camera (Hamamatsu Orca AG). Each full image of the network is assembled from images obtained over the four quadrants (Figure 3 (e2)) of the network, over a total period of ~2 s. An aqueous solution of fluorescein was utilized. Fluid flow was controlled with a syringe pump employing flow rates ranging from Q = 0.05-2 µL/min. These flow rates correspond to capillary numbers, Ca, in the range of $10^{-7}$-$10^{-8}$, which are representative of the GDL in fuel cell operating conditions. The viscosity ration M=$\mu_{water}$/$\mu_{air}$ for our case, water-to-air injection, was about 50.

**Fractal Dimension**

The fractal dimension is a measure of how the invading phase 'takes up space' within the media or network (Feder, 1988). Mathematically, the fractal dimension is defined as

$$D_F = 1 - \frac{\log(F)}{\log(L)} \quad \text{Equation 1}$$

where F is the total extent (or length) of the considered pattern, and L is the metric (or length scale) on which it is measured. In the context of a planar geometry, there are two limiting cases of fractal dimension to be considered: A line or circle has a fractal dimension of $D_F = 1$; and a solid area or disk has a fractal dimension of $D_F = 2$. For a given pattern of phase percolation within a microfluidic network, the fractal dimension is thus expected to lie in the range $\{1 < D_F < 2\}$, and may be calculated from the experimentally obtained fluorescence image. The box-counting method implemented in Benoit 1.3 was applied to calculate $D_F$ for the flow patterns. For this method, the 'yard-stick' was varied from the average channel width to 1/3 the length of the network. The fractal dimension provides a useful tool for classifying flow patterns in our pore networks. Invasion due to viscous fingering occurs when the invasion is locally non-correlated and the paths are non-overlapping; the diffusion limited aggregation model (DLA) describes this invading scenario (Halsey 2000). Invasion due to capillary fingering occurs when invading paths are still non-correlated, but experience overlap resulting in secondary phase entrapment. This behaviour corresponds to the invasion percolation with trapping model (IPT) (Wilkinson and Willemsen 1983).

**RESULTS & DISCUSSION**

**Flow patterns**

Numerical and experimental pore networks with identical network templates were invaded with liquid water. Figure 4 shows the numerically and experimentally determined breakthrough patterns for 48 x 48 pore networks with (a) isotropic perturbations and (b) diagonally biased perturbations. For each network design in Figure 4 shown from left to right are: a scaled down schematic version of the photomask, the numerical breakthrough and the experimental breakthrough patterns. Figure 4 shows that the influence of relatively minor long range perturbations can significantly alter liquid water transport in the porous medium.

By comparing the percolation patterns shown in Figure 4 (a) and (b), the effect of a diagonal bias on the pore network compared to the isotropically random pore network is evident. This influence on transport is particularly significant considering that the visual difference in network geometry is relatively minor (compare the "Photomask" column), and cannot be detected when comparing their respective photomasks alone.

**Saturation & Fractal Dimension**

By combining the total saturation, $S_t$, (measured as a ratio of the filled channels to the total number of the network channels) and the fractal dimension, $D$, we may plot a diagram consisting of our numerical and experimental data, Figure 5. This diagram demonstrates how the flow pattern evolves depending on the flow rate and time. After injection water begins filling the hydrophobic network increasing both the network saturation and the pattern fractality (Figure 5 (b)). We calculated the saturation and fractal dimension values for each flow pattern, and the result is presented on Figure 5 (b). The structures of liquid patterns corresponding to invasion are presented in Figure 5 (a). The pattern structure grows with time, as reflected in the increasing values of total network saturation, $S_t$, and the pattern fractality, D (Figure 5 (b)).

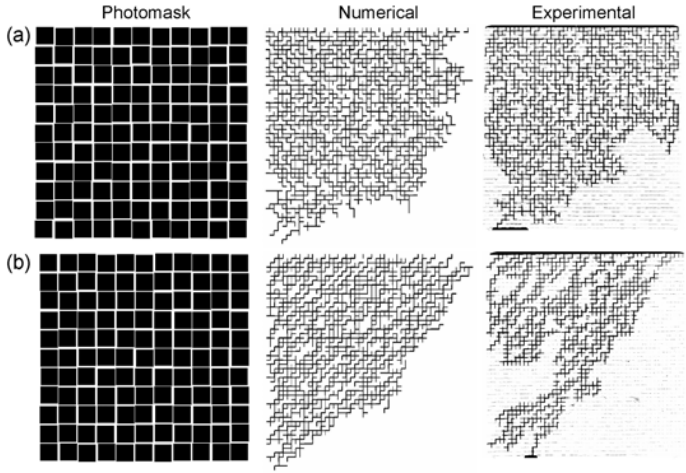

**Figure 4.** Comparison of numerical and experimental random pore network breakthrough patterns for (shown left to right): (a) isotropic perturbations, (b) diagonally biased perturbations. Each numerical and experimental result comparison is accompanied by a portion of the actual photomask schematic with a reduced number of obstacles for clarity (along left column).

If we shadow the interval of the fractal dimension between $D_{DLA}$ and $D_{IPT,}$ and reflect this area on the coordinate, $S_t$, through the curves plotted on Figure 5 (b), we will obtain an interval of the total saturations corresponding to different fractals spanning all possible patterns in the GDL. Then, it is logical to define a GDL with $S_t < S_t(DLA)$ as a dry GDL and a GDL with $S_t > S_t(IPT)$ as a wet GDL. In the context of PEM fuel cell operation, regions of the GDL near the catalyst layer and near the gas channel are expected to have varying levels of saturation. Our experiments and simulations from Figure 5 show that during invasion the structure of liquid patterns evolves from being less saturated (DLA - limit) to being highly saturated (IPT - limit). This observation is complimentary to the findings of Lenormand et al. (1988) where both the DLA and IPT regimes were structurally isolated on a (LogM, LogCa)-flow diagram. Indeed, we present a time-evolution of the flow patterns whereas Lenormand et al. (1988) reported flow diagrams for the final percolation patterns. After percolation, our patterns fall in the IPT domain in full correspondence with the (LogM, LogCa)-flow diagram reported by Lenormand et al. (1988).Therefore, this proposed method is able to span a variety of liquid pattern structures that may be of significant interest for PEM fuel cells. Particularly, the next step of developing our pore network model is to apply this method to design GDL geometries that spatially separate the DLA (dry) and IPT (wet) zones inside the network. This research would be expected to bring insight into the effect of the microporous layer, MPL, whose effectiveness has been well documented but rather poorly understood.

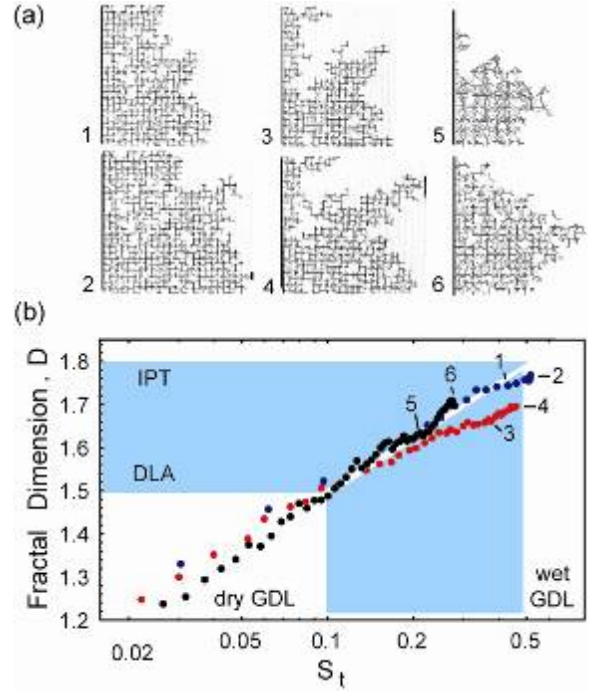

**Figure 5.** Fractal dimension versus saturation diagram in semi-log scale. (a) shows the characteristic patterns corresponding to the different flow rates of invasion, numbers denote the moments when the image of the low rate was registered. Blue, red, and black points represent the following flow conditions: 0.5 µL/min, 0.01 µL/min, and quasi-static numerical calculation. The blue zones depict an interval of parameters interesting for GDL applications; for abbreviations see the text.

**CONCLUSION**

In this work, we discussed the first stage of developing and validating a GDL-representative pore network model. We began with a two-dimensional pore network model with a single mobile phase invading a hydrophobic media, whereby the slow capillary dominated flow process followed invasion percolation. The procedure to design isotropically and diagonally biased networks was described, and the experimental techniques employed to provide feedback to the model were also described. Microfluidic network fabrication using soft lithography was described, and the experimental and numerical flow patterns showed reasonable agreement. To further characterize the experimental and numerically predicted flow patterns, the fractal dimension and saturation were introduced as a promising method to classify PEM fuel cell GDL operating regimes in terms of the well-documented porous media flow regimes (DLA and IPT). Furthermore, future work that includes employing this classification method to design GDL geometries that resemble the behaviour of the microporous layer.

**ACKNOWLEDGMENTS**

The authors are grateful for the financial support of the Natural Sciences and Engineering Research Council (NSERC) of Canada, the Canada Research Chair Program, and the National Research Council (NRC). The authors would also like to acknowledge the support of the Canadian Foundation for Innovation (CFI).